\documentclass[preprint]{revtex4}
\usepackage{graphicx}
\usepackage{bm}
\begin{document}
\title{Strongly coupled modes in a weakly driven micromechanical resonator}
\author{W.J. Venstra}
\author{R. van Leeuwen}
\author{H.S.J. van der Zant}
\affiliation{Kavli Institute of Nanoscience, Delft University of Technology, Lorentzweg 1, 2628CJ Delft, The Netherlands}

\begin{abstract}
We demonstrate strong coupling between the flexural vibration modes of a clamped-clamped micromechanical resonator vibrating at low amplitudes. This coupling enables the direct measurement of the frequency response via amplitude- and phase modulation schemes using the fundamental mode as a mechanical detector. In the linear regime, a frequency shift of $\mathrm{0.8\,Hz}$ is observed for a mode with a line width of $\mathrm{5.8\,Hz}$ in vacuum. The measured response is well-described by the analytical model based on the Euler-Bernoulli beam including tension. Calculations predict an upper limit for the room-temperature  Q-factor of $\mathrm{4.5\times10^5}$ for our top-down fabricated micromechanical beam resonators.
\end{abstract}
\maketitle
Nonlinear interactions between the vibration modes in micro- and nanomechanical resonators have attracted significant interest recently. In extensional structures, such as clamped-clamped bridges, the modes are coupled by the displacement-induced tension~\cite{Westra10}, which yields a quadratic relation between the resonance frequency of the mode considered and the amplitudes of the other modes. In singly-clamped cantilevers the displacement-induced tension is absent; here the inextensionality condition couples the horizontal and vertical displacements of all modes, resulting in qualitatively similar dynamics~\cite{Crespo78b,Kacem10,Venstra10,Westra12}. Several applications and consequences have been put forward based on the modal interactions, such as enhancement of the dynamic range~\cite{Westra10}, modification of the resonator damping by employing phonon-phonon cavities~\cite{Venstra11, Mahboob12}, frequency stabilization~\cite{Antonio12}, the study of relaxation mechanisms~\cite{Faust12} and linear frequency conversion~\cite{Westra11}. In thermal equilibrium, via the modal interactions the displacement fluctuations in one mode give rise to  frequency fluctuations in the other modes, thus broadening their spectral line. In recent theoretic work these frequency fluctuations were quantified for a carbon nanotube~\cite{Barnard11}, yielding a boundary for the experimental Q-factor of such resonators at room temperature. Experiments on suspended carbon nanotube resonators in the Coulomb blockade regime demonstrate that single-electron-tunneling processes provide a strong electrostatic coupling between the modes, in addition to the mechanical mode coupling~\cite{Castellanos12, Eichler12}.\\
\begin{figure}[b]
\includegraphics[width=85mm]{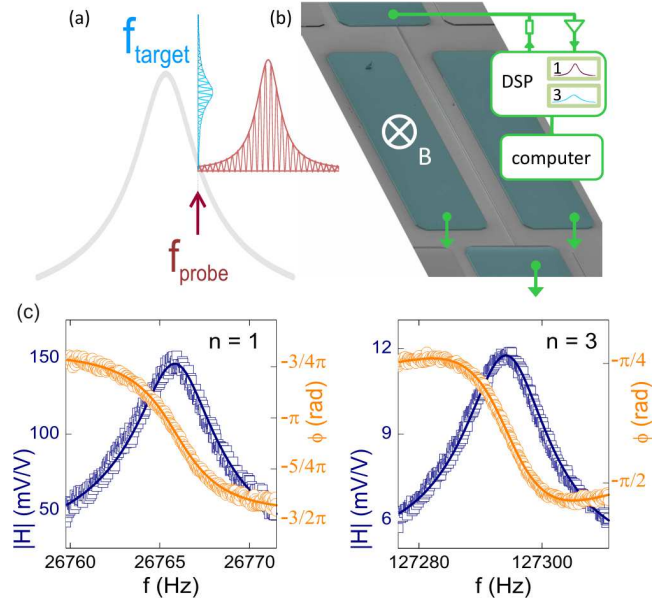}
\caption{(a) Amplitude modulation scheme to probe the modal interactions at the low amplitudes. The response of the probe mode is modulated by the displacement-induced tension of the target mode. (b) Measurement setup and colored scanning electron micrograph (top view) of the silicon (110) beam resonator. The motion is in the plane of the figure. (c) Linear frequency response of the probe and target modes.}
\end{figure}
\indent\indent While the recent experimental work has focused on the mode coupling in strongly driven resonators exhibiting nonlinear vibrations, in this Letter we investigate these interactions in the low-amplitude regime. We demonstrate that the modal interactions play a significant role in the dynamic behavior of a linear resonator, as the vibrations of a weakly driven mode modulate the motion of a second vibration mode. We employ this coupling to perform swept-frequency type measurements of the linear frequency response of a high frequency (target) mode, by measuring the induced amplitude (AM) and phase modulation (PM) in a low frequency (probe) mode which is weakly driven at a fixed frequency. This provides a practical way to measure the frequency response, and it should be contrasted to the scheme presented earlier~\cite{Westra10}, where the frequency response of the target mode is reconstructed from a series of resonance frequency measurements on the probe mode. The observed modulation depth is in agreement with a model based on the Euler Bernoulli beam including tension~\cite{Westra10}. Using this model with the parameters extracted from experiment, the frequency broadening of the fundamental mode that occur when the high amplitude mode is in thermal equilibrium is estimated. An upper limit $\mathrm{Q=4.5\cdot 10^5}$ is found for our  micrometer-scale silicon beam resonator. This indicates that modal interactions play a significant role in the experimentally observed Q-factors of micromechanical resonators.\\
\indent\indent To detect the modal interactions in the linear regime we deploy the fundamental flexural mode as the probe, by driving it at a fixed frequency $\mathrm{f_{probe}}$. Its response is modulated by the averaged tension induced by the displacement of a second mode of the resonator, which is driven at $\mathrm{f_{target}}$. Fig. 1(a) shows the principle of this AM scheme. The experiments are conducted using silicon beams with dimensions $\mathrm{L\times w \times h = 1000 \times 2\times 2 \,\mu m^3}$, fabricated by anisotropic wet etching of silicon-on-insulator wafers with an (110)-oriented device layer~\cite{Leeuwen11}. Fig.1(b) shows the resonator and the setup; the beam is driven by applying an alternating current in the presence of a permanent magnetic field~\cite{Venstra09}, via a thin conductive layer evaporated on top (70 nm of chromium/gold). The driving force and the detector couple to the in-plane motion of the device. To eliminate broadening of the resonance peak by the viscous force, the experiment is conducted at a pressure of $\mathrm{\approx 10^{-4}\,mbar}$. Frequency response measurements when one mode is driven and the other mode is in thermal equilibrium are shown in Fig. 1(c). The linear resonance frequencies for the first and third mode ($\mathrm{n=1,3}$), $\mathrm{f_{R,1}=26765.7\,Hz}$ and $\mathrm{f_{R,3}=127294\,Hz}$, and the corresponding Q-factors, $\mathrm{Q_1=4527}$ and $\mathrm{Q_3=7476}$, are obtained from damped-driven harmonic oscillator fits (solid lines)~\cite{stress}.\\
\indent\indent The sensitivity of the probe mode to the motion of the target mode varies with $\mathrm{f_{probe}}$ and maximizes when the second derivative of the probe mode frequency response is zero. In Fig. 2(a) the fitted amplitude (left) and phase (right) responses of the fundamental mode (grey) of Figure 1 are repeated, together with their second derivatives (blue solid lines). The inflection points, marked B and D, occur at detuning $\mathrm{\Delta f\approx \frac{1}{Q\sqrt{8}}}$, whereas the phase modulation maximizes on resonance, $\mathrm{\Delta f=0}$ (C).  Figure 2(b) shows measurements of the magnitude (left) and phase (right) response of the probe mode while sweeping $\mathrm{f_{target}}$ through the resonance frequency of mode 3. The (fixed) drive frequencies correspond to A-E in panel (a). At $\mathrm{f_{probe}=26.76\,kHz}$, (A), the probe mode is off-resonance and the target mode is not detected. In B, on the positive slope of the frequency response curve of the probe mode, the amplitude modulation maximizes. The tension of the target mode tunes the resonance frequency of the probe mode to a higher value, and produces in a dip in the amplitude response of the probe mode. When the probe mode is on resonance (C) the amplitude modulation is close to zero: here the phase response maximizes. Probing at a higher frequency results in maximization of the AM response in D, and off-resonance (E)  the modulation is zero. These measurements demonstrate that the fundamental mode can be conveniently used to probe the response of a mode that is 5 times higher in frequency, in a swept-frequency type measurement.\\
\begin{figure}[b]
\includegraphics[width=70mm]{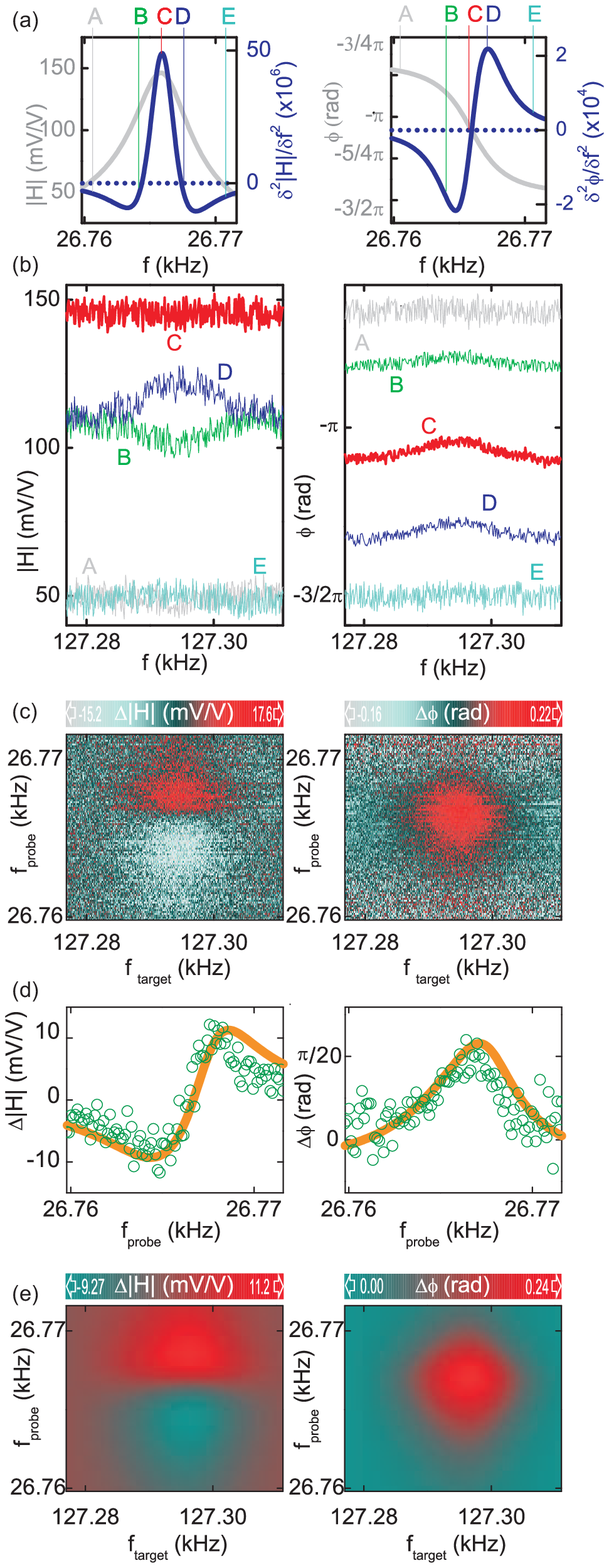}
\caption{(a) Frequency response of the probe (fundamental) mode and its second derivative. (b) Amplitude (left) and phase (right) response of the probe mode driven at the frequencies indicated in (a), while driving the target mode through resonance.(c) Amplitude (color scale, left) and phase (right) of the probe mode, after subtracting the harmonic oscillator response background. (d) Sensitivity of the probe mode in AM and PM modulation schemes as a function of the drive frequency of the probe (solid lines: calculation). The target mode is driven on resonance. (e) Response as calculated by solving the coupled equations of motion (see text).}
\end{figure}
\indent\indent Figure  2(c) shows the full measurement from which the traces are extracted. The target mode is swept along the horizontal (fast) axis, while the frequency of the probe mode is incremented along the vertical (slow) axis. The color scales represent the amplitude (left) and phase (right) response of the probe mode. The probe response when the target mode is off-resonance is subtracted to reveal the resonant features more clearly. Figure 2(d) shows the modulation depth as a function of the probe mode frequency, when the target mode is driven on-resonance. The AM modulation depth is $\mathrm{10\,mV}$, which corresponds to a frequency tuning of $\mathrm{0.8\,Hz}$. This frequency shift is of the same order as the line-width of the resonator, $\mathrm{5.8\,Hz}$, indicating the strong coupling between the modes vibrating in the linear regime. Notably, in these measurements the applied drive force for both modes is the same as in Fig. 1(c), and no apparent sign of nonlinearity is observed in the line shapes. In Fig. 2 (d) the back-action from the target mode causes a small but noticeable upwards shift of the resonance frequency of the probe mode response.\\
\indent\indent To describe the modulation between the two linear driven modes, we calculate the complex response of mode 1 while driving mode 3 through resonance, using the model developed in Ref.~\cite{Westra10}. The amplitudes $\mathrm{a_{1,3}}$ of modes 1,3 driven at frequency $\mathrm{\omega_{1,3}}$ are then given by:
\begin{eqnarray}
\Big( \omega_{R,1}^2- \omega_1^2 + i\omega_1\omega_{R,1}/Q_1+ \frac{\tau}{4}|a_1|^2 I_{11}^2 + \frac{\tau}{4}\Big(|a_3|^2I_{11}I_{33} +|a_3|^2I_{13}^2 \Big)\Big)a_1 = \int_0^1 f_{\mathrm{ac,1}} \xi_1 \mathrm{d}x,\\
\Big( \omega_{R,3}^2- \omega_3^2 + i\omega_3\omega_{R,3}/Q_3+ \frac{\tau}{4}|a_3|^2 I_{33}^2 + \frac{\tau}{4}\Big(|a_1|^2 I_{33}I_{11}+|a_1|^2 I{31}^2\Big)\Big)a_3= \int_0^1 f_{\mathrm{ac,3}} \xi_3 \mathrm{d}x.
\end{eqnarray}
The scaled Lorentz force per unit length acting on the beam is $f_{\mathrm{ac},i}= \frac{12L^4}{Ewh^4}B I_i$, where  $I_1=715\,\mathrm{pA}$ and $I_3=10.7\,\mathrm{nA}$ are the rms currents through the resonator and $B=1.9\,T$ is the permanent magnetic field. For a
stress-free beam with a rectangular cross section $\tau=12$, the integrals $I_{ij}=\int_0^1 \xi_i'(x)\xi_j'(x)\mathrm{dx}$ are $I_{11} = 12.3$, $I_{33} = 98.9$ and $_{13}=I_{31}=-9.7$, and the average displacements per unit deflection are $\int^1_0 \xi_1dx'=0.83$
and $\int^1_0 \xi_3 dx'=0.36$ for the beam-like mode shapes $\xi_i$. Young's modulus equals $\mathrm{E=169\,GPa}$ for our device, which bends about the silicon (110) axis~\cite{Hopcroft10}. Figure 2(d) (solid lines) shows the modulation depth as calculated by numerically solving the coupled equations of motion. Agreement is found between the measured and the calculated sensitivity curves, where the phase shift is quantitatively predicted by the model, and the  magnitude is scaled by a constant factor as the detector gain is not accurately known in this experiment.\\
\indent\indent The above analysis demonstrates that the modal interactions give rise to a significant tuning of the modes in the low-amplitude regime. The driven displacements of mode 1 in the experiment exceed the thermomechanical fluctuations by only 2 orders of magnitude, and introduce a significant frequency tuning when compared to the resonator bandwidth. It is interesting to consider the frequency shifts when the modes are in thermal equilibrium and the external driving force is zero. Here, the displacement fluctuations give rise to frequency fluctuations via the displacement-induced tension. For carbon nanotube resonators this broadening was calculated using a thermodynamic approach and a discretized model of the resonator~\cite{Barnard11}. With the continuous Euler-Bernoulli theory, which accurately describes the experimentally observed interaction between the modes at low amplitudes, the frequency shift for our top-down fabricated micromechanical resonators equals $\mathrm{0.09\,Hz/nm^2}$. For an rms thermal displacement $\mathrm{a_{3, th}=\sqrt{k_BT/m\omega_{R,3}^2}=0.026\,nm}$ this figure implies an upper bound for the experimentally observed Q-factor of mode 1 of approximately $\mathrm{ Q_{1,max}<4.5\cdot10^5}$ at room temperature, due to the interaction with mode 3. This value will be lower when the motion of other degrees of freedom are taken into account.\\
\indent\indent The predicted upper bound is lower than the experimental Q-factors for string resonators under tension, for which values over a million have been reported at room-temperature~\cite{Verbridge08}. The difference can be explained as follows. The string devices incorporate a large residual tension, which forms the main contribution to the restoring force~\cite{Unterreithmeier10}. As a result, the flexural displacements due to thermal energy are small, and so is the displacement-induced tension. The dispersion due to a fluctuating displacement via the modal interactions therefore reduces with the residual tension. This is mathematically expressed via the interaction matrix $I_{ij}$, which for devices with negligible bending rigidity (strings) contains only nonzero diagonal terms ($\mathrm{I_{i\neq j}=0}$), thus limiting the available dispersive interactions for strings. Finally, in a perfect string the spectrum is harmonic with $\mathrm{f_{R,n}=nf_{R,1}}$. In this case the energy is recycled between the modes as the motion of one mode gives rise to a parametric excitation of the other modes via the displacement-induced tension. In a perfect string this parametric excitation is effective, as it always occurs at integer multiples of the resonance frequencies and all harmonics are parametrically degenerate.\\
\indent\indent In conclusion, we have studied the modal interactions in clamped-clamped resonators in the regime of linear vibrations. At drive strengths that are 5 orders of magnitude lower than in previous experiments on micromechanical resonators~\cite{Westra10}, we observe a strong mechanical interaction between the modes, inducing frequency shifts on the order of the resonance line width in vacuum. The interaction is employed to perform swept-frequency type measurements by measuring the amplitude and phase modulation of the fundamental by a mode that is 5 times higher in frequency. The presented scheme, i.e. coupling the motion to a mechanical detector, may be favorable compared to an electronic detector as it does, e.g., not suffer from signal loss due to parasitic capacitance. The amplitude and phase modulation is quantitatively described by an Euler-Bernoulli beam model including tension. This model predicts that the fluctuating displacements in thermal equilibrium result in frequency broadening, and limit the room-temperature Q-factor of our micromechanical resonators to $\mathrm{4.5\cdot 10^5}$. In reality this value will be lower since the model only considers broadening by the third mode, and neglects the contribution of the other modes.\\
\indent\indent This work is supported by FOM (Program 10, Physics for Technology) and NanoNextNL, a micro- and nanotechnology consortium of the Government of the Netherlands and 130 partners.

\bibliographystyle{apsrev}

\begin{thebibliography}{19}
\bibitem[{str()}]{stress}
\bibinfo{note}{The ratio $\mathrm{f_3/f_1=4.8}$ is slightly lower than $5.4$
  for a stress-free beam resonator, indicating a small amount of tensile
  residual stress. This stress is introduced during backside thinning of the
  wafer, and has no consequences for the further analysis.}
\end{thebibliography}

\end{document}